\documentclass[traditabstract]{aa} % for a referee version add ,referee
\usepackage{txfonts,psfig}
\usepackage{color}

\begin{document}

\title{Relative distribution of dark matter and stellar mass in three
massive galaxy clusters}
\titlerunning{Stellar-to-dark matter mass radial profiles} 
\author{S. Andreon\inst{1}}
\authorrunning{S. Andreon}
\institute{
INAF--Osservatorio Astronomico di Brera, via Brera 28, 20121, Milano, Italy,
\email{stefano.andreon@brera.inaf.it} \\
}
\date{Accepted ... Received ...}
\abstract{
This work observationally addresses the relative
distribution of total and
optically luminous matter in galaxy clusters
by computing the
radial profile of the stellar-to-total mass ratio. We adopt
state-of-the-art accurate lensing masses free
from assumptions about the mass radial profile and we use
extremely deep 
multicolor wide--field optical images to distinguish star formation 
from stellar mass, to properly calculate the mass 
in galaxies of low mass, those
outside the red sequence, and to allow a contribution from
galaxies of low mass that is clustercentric dependent. We pay special
attention to issues and contributions that are usually underrated, 
yet are major sources of uncertainty,
and we present an approach that allows us to account for all of them.
Here we present the results for three
very massive clusters at $z\sim0.45$, MACSJ1206.2-0847, 
MACSJ0329.6-0211, and RXJ1347.5-1145. We find
that stellar mass and total matter are closely distributed on scales
from about 150 kpc to 2.5 Mpc: the stellar-to-total mass ratio
is radially constant. We find that the characteristic mass stays 
constant across clustercentric radii and clusters,
but that the less-massive
end of the galaxy mass function is dependent on the environment.
}

\maketitle

\section{Introduction}

Clusters of galaxies are the largest dynamically bound systems and
are effective laboratories for studying the relationship between
the distribution of dark and luminous matter. 
Since the pioneering work of Zwicky (1933), many studies
(e.g., Fabricant et al. 1986; Eyles et al. 1991;  
Smail et al. 1995; Kaiser et al. 1998;
Kneib et al. 2003; Medezinski et al. 2010; Okabe et al. 2014;
Grillo et al. 2014) have analyzed the relative distribution
of luminous and dark matter mass using the mass--over--luminosity (M/L) or 
its reciprocal, light--over--mass (L/M) ratio.
Zwicky (1933) provided the first evidence
of the existence of  dark matter. Most weak--lensing studies, and several
X--ray studies, find that the dark--matter 
component is well traced by the total light distribution of the cluster
(e.g., Fabricant et al. 1986; Smail et al. 1995;
Kneib et al. 2003; Medezinsky et al. 2013; Okabe et al. 2014
Grillo et al. 2014). We note, however, that some lensing studies 
{\em assume} that the above holds. In fact, they take
the light distribution as prior of the
mass distribution, for example, holding the center of the sub--halos tied
on the galaxy centers and scaling their mass by the galaxy luminosity
(e.g., Grillo et al. 2014, Kneib et al. 1996).  

Quantitatively, we are far from a consensus.
Smail et al. (1995) and Annunziatella et al. (2014) find some evidence for
a greater concentration of total matter compared to galaxies.
Medezinsky et al. (2013)
find a radially decreasing $M/L$ outside the cluster core,
as Rines et al. (2000) for  Abell 576 using caustics.
Rines et al. (2004) find a decrease by a factor of two
outside the virial radius  in a combined sample of
nine clusters including Coma and Abell 576. Instead, 
Hoekstra et al. (1998), Kneib et al. (2003), and others, 
measure a constant $M/L$ with radius. Similarly, Rines et al. (2001) find, 
using caustics to estimate mass (e.g., Diaferio \& Geller 1997),
that the Coma cluster has a flat $M/L$ over all scales going from the
cluster center to several virial radii.  At the cluster's very center,
Newman et al. (2013) find that within the brightest
cluster galaxy effective radius almost all of the matter is
formed by stars, i.e., that the stellar fraction is close to one at $r=0$. 
Instead, Grillo et al. (2014) found a vanishing amount
of stellar mass at the cluster center for the single cluster they studied. We note, however, that 
while Newman et al. (2013) take the cluster center co-incident with the center 
of a galaxy (the brightest one), Grillo et al. (2014)
adopt a cluster center away from any galaxy, making the variance between the
two results somewhat expected. 
Finally, Andreon (2012) measured  the stellar masses for 12 clusters within
two radii, $r_{500}$ and $r_{200}$, finding
$\log M^{stars}_{200}-\log M^{stars}_{500} \sim 0.12$ dex, close
to the measured difference in mass between these radii. 
However, the masses at the largest radii are based on
extrapolation because X--ray data do not
reach that far.

Recent 
hydrodynamical simulations (Battaglia et al. 2013, Planelles et al.
2013) implementing radiative cooling, star formation,
and AGN feedback predict that stars in clusters mirror the
dark matter distribution, i.e., that the stellar-to-total
mass ratio is quite constant across most of the cluster radial range,
with the possible exception of radii very close to the cluster center, where 
predictions are less accurate (see Newman et al. 2013). 
The constancy of the radial profile
of the stellar--to--total-matter
mass ratio agrees with qualitative expectations:
stars are not expected to be stripped significantly because they are located
deep in the galaxy potential, hence stellar mass should trace
dark mass given that both are collisionless. However,
from a theoretical and quantitative perspective, we are 
far from a satisfactory understanding 
because current
numerical simulations grossly overpredict the total amount
of stellar mass (e.g., Andreon 2010; Young et al. 2011; 
Planelles et al. 2013).
 
Impressive progress on the measurements of 
stellar mass and total cluster luminosity 
has occurred in  recent years, overcoming
limitations of past works, most of which, for example, 
did not consider 
galaxies outside the red sequence, or even in it, but
in galaxies of low mass. These stellar masses are missed most of
the time, or strong assumptions have been made  
about them, such as assuming a fixed proportion
of massive and less massive galaxies. 
Some works assumed that the luminosity
function is radial--independent, while  luminosity
segregation is known to exist in clusters (e.g., Andreon 2002b; Durret et al. 2002).  
Works using early releases
of the Two Micron All Sky Survey (Skrutskie et al. 2006) photometry 
are missing light from outer regions
of the detected galaxies and even whole galaxies of central
surface brightness below average because of the shallow
nature of the used data (e.g., Andreon 2002a; Kirby et al. 2008).
In some rare cases, researchers were presented with
data only allowing a coarse, or no, background subtraction.

There  has also been impressive progress on
the determination of the total matter radial profile, whose
determination is hampered
by substructure and deviations of clusters from 
equilibrium, which affect both X--ray observations and Jeans
analyses (i.e., kinematics). Indeed, kinematic
analysis assumes a) dynamical equilibrium, b) some
dynamical modeling (on the distribution of orbits, for example), 
and c) knowledge of the velocity 
bias of the observed galaxy population. Stacking clusters erase 
the signature of the substructure on the
combined sample, but the individual clusters  in the sample
are not for this reason closer to equilibrium. Removing  galaxies in
substructures from
the stack is of no help in removing their effect on the other galaxies.
X--ray estimates very rarely reach large cluster scales
because the surface brightness
emission falls precipitously with radius from the cluster
center and spatially resolved temperature profiles at large clustercentric
radii are therefore challenging. Progress has been  made quite recently (e.g., Mamon et al. 2013
for kinematical estimates and Hoshino et al. 2010 for X--ray determinations),
but early studies did not profit from this, of course.

Lensing (e.g., Mellier 1999) allows us to determine
the projected mass profile free of assumption
about (dynamical, hydrostatic, etc.) equilibrium. 
However, it is a new technique,
which has impressively improved  in various aspects over the years
(modeling of PSF effects, estimate of the redshift distribution of
the lensed sources, treatment of dilution effects, etc.), to such a degree
that accurate mass radial profiles have only recently  started
to become available  (Applegate et al. 2014; 
Umetsu et al. 2012, 2014). Furthermore, most of the mass 
profiles, including recent ones, are derived assuming a) azymuthal symmetry, 
b) a single or a double parametric 
functional form for the radial  profile, and c) the assumption
that the cluster is embedded in a smooth large--scale environment (e.g., 
Hoekstra et al. 2012; Merten
et al. 2014; von der Linden et al. 2014). While numerical simulations
predict a universal dark matter profile for the average halo 
close to a Navarro, Frenk \& White (1997, hereafter NFW) profile, 
real massive
clusters studied in weak--lensing works (and many of those
studied using caustics or kinematics) are rarely  that simple 
(e.g., Smail et al. 1995; Kneib et al. 2003; Umetsu 2013;  Umetsu et al. 2014;
Gruen et al. 2014; Klypin et al. 2014). 
Parametric estimates that assume a radial profile model that does not
fit the data (e.g., because of cluster substructure) induce biases in the 
derived $M/L$ (or stellar--to--total matter mass) ratio. In fact,  
deviations from the assumed radial model 
are not captured by the too simple model. Instead, a more flexible
weak lensing model, such as the radial model--free modeling
by Umetsu (2013) and Umetsu et al. (2014), does  capture
the  complex morphology of the cluster in a far better way.

In this work we make use of the impressive
progress in weak lensing and stellar mass measurements to
present the first robust determination
of the radial profile of the
stellar--to--total matter mass ratio. In particular,
we measure it for three massive clusters at intermediate
redshift up to the virial radius, $r_{vir}$, and beyond. 
We use up--to--date, robust, and radial--model free, 
weak lensing masses that are  also free from assumptions about
the cluster dynamical status. 
Thanks to Suprime--Cam multicolor
wide--field optical images of exquisite depth and,
unlike  most previous works using the $M/L$ ratio
as proxy for the stellar--to--total matter 
ratio, we 
distinguish luminosity emitted by 
a recent minor (in mass) burst of star formation 
from the one emitted by a much larger amount of old stellar mass;
we  count the mass in galaxies outside the red sequence 
or in galaxies of low mass; and we  allow (and we will indeed find)
a cluster--centric dependent contribution from
galaxies of low mass. Finally, we study the differential
radial profile (i.e., $f(r)$) more
sensitive to gradients than the integrated profiles (i.e., $f(<r)$)  studied
in past works. To achieve our aim, we are obliged
to first derive (and use) the correct expression of the likelihood and
to account for covariance across mass radial bins.

\begin{figure*}
\centerline{
\psfig{figure=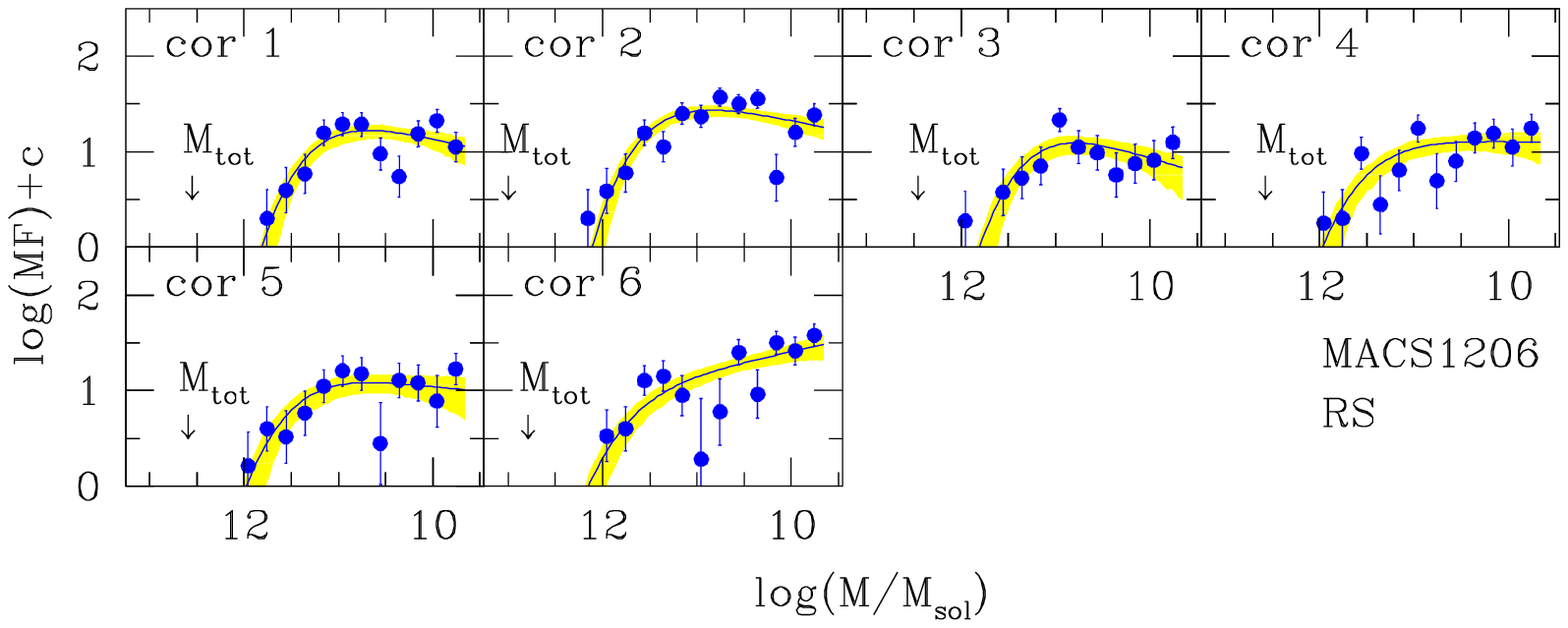,width=14truecm,clip=}
}
\centerline{
\psfig{figure=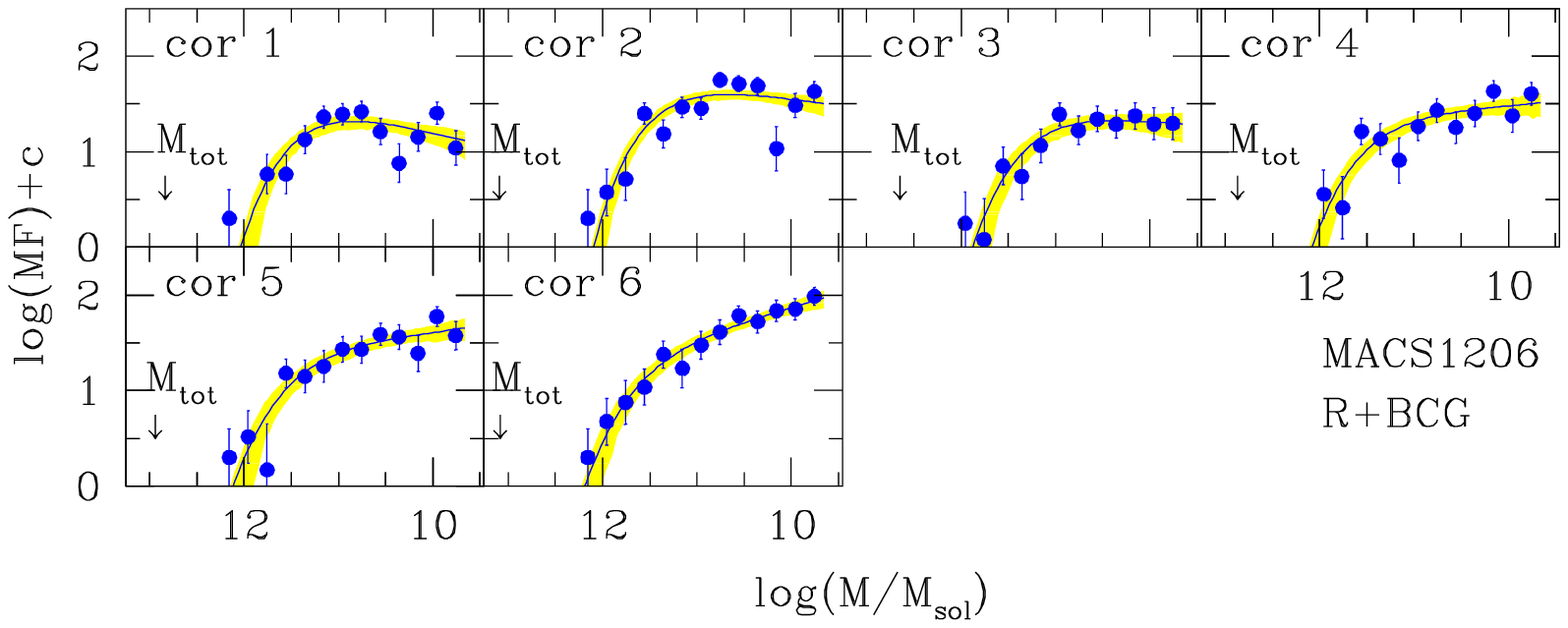,width=14truecm,clip=}
}
\caption[h]{Mass function of red sequence (upper panels) and 
red galaxies plus BCG (lower panels) of MACSJ1206 in various 
clustercentric radial bins, as detailed in the panels. The
arrow indicates the total mass inside the considered radial bin,
disregarding the BCG, if any. The solid line
is the mean model fitted to the individual galaxy data, 
while the shading indicates the 68\% uncertainty
(highest posterior density interval). Points and approximated
error bars are derived 
by binning galaxies in magnitude bins, 
adopting approximated Poisson   
errors summed in quadrature, as commonly done in the literature.
}
\end{figure*}

Throughout this paper, we assume $\Omega_M=0.3$, $\Omega_\Lambda=0.7$, 
and $H_0=70$ km s$^{-1}$ Mpc$^{-1}$. Magnitudes are in the AB system.
We use the 2003 version of Bruzual \& Charlot (2003, hereafter BC03) stellar 
population synthesis
models with solar metallicity and a Salpeter initial 
mass function (IMF). 
We define stellar mass as the integral of the star formation
rate, which includes the mass of gas processed by stars 
and returned to the interstellar medium. We allow both single
stellar populations and exponential declining $\tau$ models.
For star forming galaxies,
we also include the stellar mass plausibly formed from the redshift
of observation to $z=0$ (i.e., we extend the integral of
the star formation rate down to $z=0$).
Results of stochastic computations are given
in the form $x\pm y$, where $x$ and $y$ are 
the posterior mean and standard deviation. The latter also
corresponds to 68 \% intervals because we only summarized
posteriors close to Gaussian in this way.

\section{Samples and data}

We study three
clusters:  MACSJ1206.2-0847 ($z=0.439$), MACSJ0329.6-0211 ($z=0.450$),
and RXJ1347.5-1145 ($z=0.451$, hereafter referred to as  MACSJ1206, MACSJ0329,
and RXJ1347, respectively. These are selected among CLASH clusters (Postman
et al. 2012) with 1) state--of--the--art accurate weak lensing masses free
from assumptions about the mass radial profile (taken from Umetsu
et al. 2014), 2) with available 
Suprime--Cam images (Miyazaki et al. 2002), and 3) at $z>0.4$ in order to probe
$\gg 3$ Mpc radii at the cluster redshift (needed to estimate the background inside the same Suprime--Cam pointing).
 
The basic data used in our analysis are $V$, $R$, and $z'$ photometry
derived using the SExtractor code
(Bertin \& Arnouts 1996) on deep Subaru 
Suprime--Cam  images delivered by the
CLASH team\footnote{http://archive.stsci.edu/prepds/clash/}, described
in Umetsu et al. (2014; see also Medezinski et al. 2013). 
Suprime--Cam images are $\approx 25'\times25'$ wide, 
have a pixel size of $0.2$ arcsec, and are taken in sub--arcsec seeing
conditions. 
We emphasize that the Suprime--Cam filters differ
from the standard $V$, $R$, and $z'$ filters. We therefore compute
the combined response of the filters, prime focus unit transmission,
and CCD quantum efficiency\footnote{http://www.naoj.org/Observing/Instruments/SCam/} and used these for the
stellar population synthesis modeling. These filters,
at $z=0.44$, accurately sample the 4000 \AA \ break, 
approximately corresponding to the $u,B,$ and $R$ filters. 
These are among the deepest cluster observations ever taken
(e.g., Medezinski et al. 2010).  

We conservatively consider
galaxies brighter than $z'=24$ mag, which is much
much brighter than the completeness 
limit of these images 
(e.g., by $2$ mag in $z'$; see, e.g., Medezinski et al. 2010),
to ignore incompleteness and its radial dependency
(due to dithering, CCD gaps, variable source density or crowding,
etc.).
We adopt SExtractor's magauto as total galaxy magnitudes, while colors
are based on a fixed 2 arcsec aperture and are
corrected for the color--magnitude 
slope.
Since early-type galaxies have surface brightness
profiles close enough to make global properties such as total luminosity
independent of environment (e.g., Pahre et al. 1998), we do not
expect that the precise definition of total mag affects
the major conclusion of this work, namely a radially
constant stellar--to--total mass ratio.

Two galaxy populations are of interest:
\begin{itemize}
\item red sequence (RS) galaxies are defined as those within 0.1 mag redward
and 0.2 mag blueward in $V-z'$ with respect to the red sequence;

\item red galaxies plus brightest cluster galaxies (R+BCGs). The former are 
defined as those redder than an exponential
declining $\tau=3.7$ Gyr model, and bluer than 0.1 mag redward of 
the color--magnitude relation, as in several
previous studies (e.g., Andreon et al. 2006; Raichoor \& Andreon 2012a,b, Andreon 2012).
These galaxies will be red  at $z=0$ according to the Butcher \& Oemler
(1984) definition. Of course, this sample includes RS galaxies.
In the inner cluster radius a sizeable amount
of stellar mass is missing for clusters with blue BCGs
(one of the two BCGs of both MACSJ0329 and RXJ1347). 
Therefore, we add these two BCGs to the red galaxy population.

\end{itemize}

We also consider other, bluer, galaxy populations. However, we found that
they are only  minor populations   for the  three clusters studied
and, as  populations, marginally
detected, if at all, in the background at most or all clustercentric radii,
in agreement with the lack of a blue population in Andreon et al. (2006), 
De Propris et al. (2003), Haines et al. (2009), and Raichoor \& Andreon (2012b).
Given that in the studied clusters the population is small and of low mass for their
luminosity --they are bright because they are strongly star forming, not 
because they are massive-- we can safely
neglect them in our stellar mass computation. Indeed, results for the R+BCG
sample turns out not to depend  on the details of the red definition (for
example by adopting a $\tau=5$ model)  because the adopted color boundaries
fall (by design) in regions where no cluster galaxies (in an amount large
enough to be detected in the background) are found. Using the
large sample of spectroscopic redshift of MACS1206 (Biviano et al. 2013),
which is biased against red galaxies,
we found that less than 3\% of the stellar mass is in bluer galaxies.

We  estimate galaxy stellar masses from the $z'$ band luminosity
and $V-z'$ color using the relation between $M/L$ and galaxy color
derived using the  Bruzual \& Charlot (2003) stellar population synthesis models
(like that of Bell \& de Jong 2001, and similarly to Kravtsov et al. (2014),
Boselli et al. (2014), Fritz et al. (2014), Budzynski et al. (2014), 
Gladders et al. (2013), Oemler et al. (2013), and Dressler et al. (2013),
to only mention cluster works that have appeared in the last year. 
This single--color estimate of stellar
mass is found to give values consistent with full--SED fitting
up to $z\sim2$ (Koyama et al. 2013).
In particular, the slope of $M/L$ vs $V-z'$ relation
is derived by computing the color
(in the observer frame for galaxies at the cluster redshift) 
of $0<\tau<5$ Gyr model galaxies.

\begin{figure*}
\centerline{
\psfig{figure=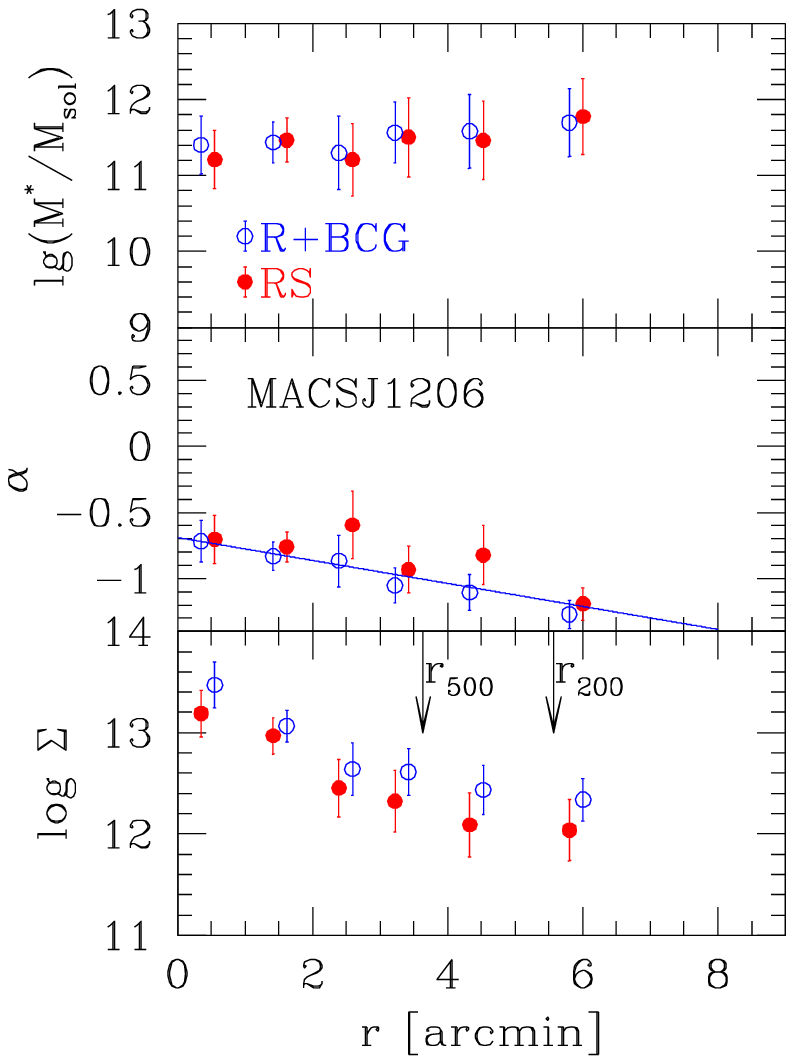,width=6truecm,clip=}
\psfig{figure=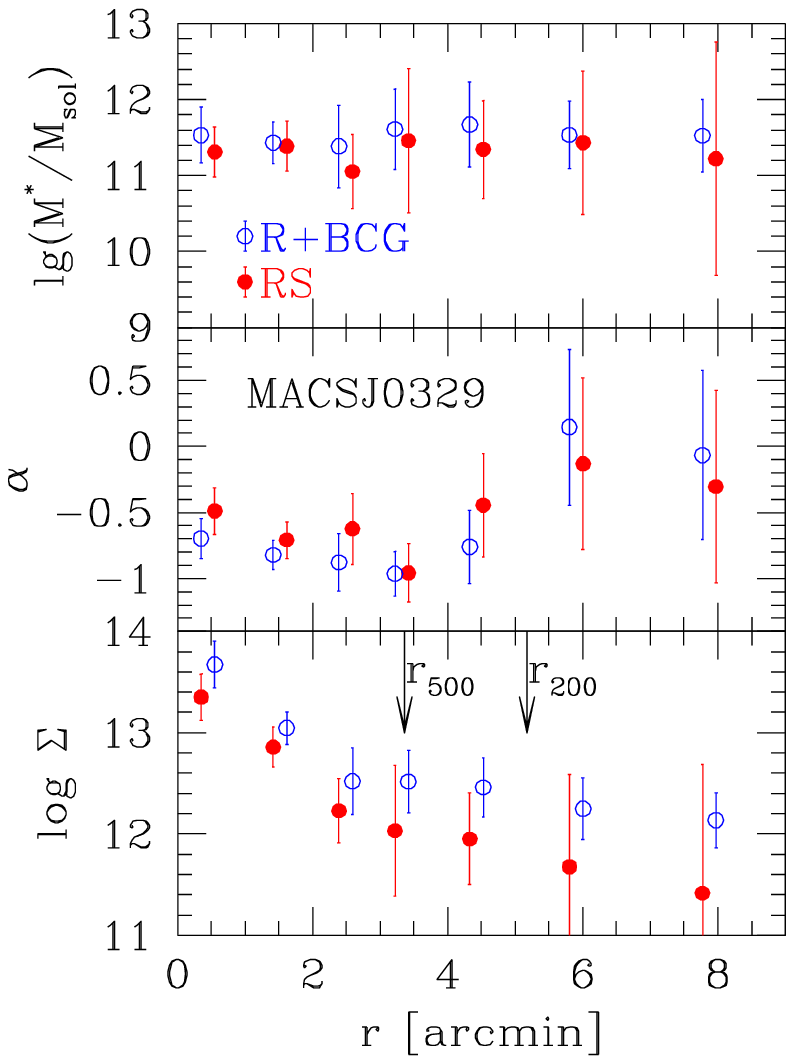,width=6truecm,clip=}
\psfig{figure=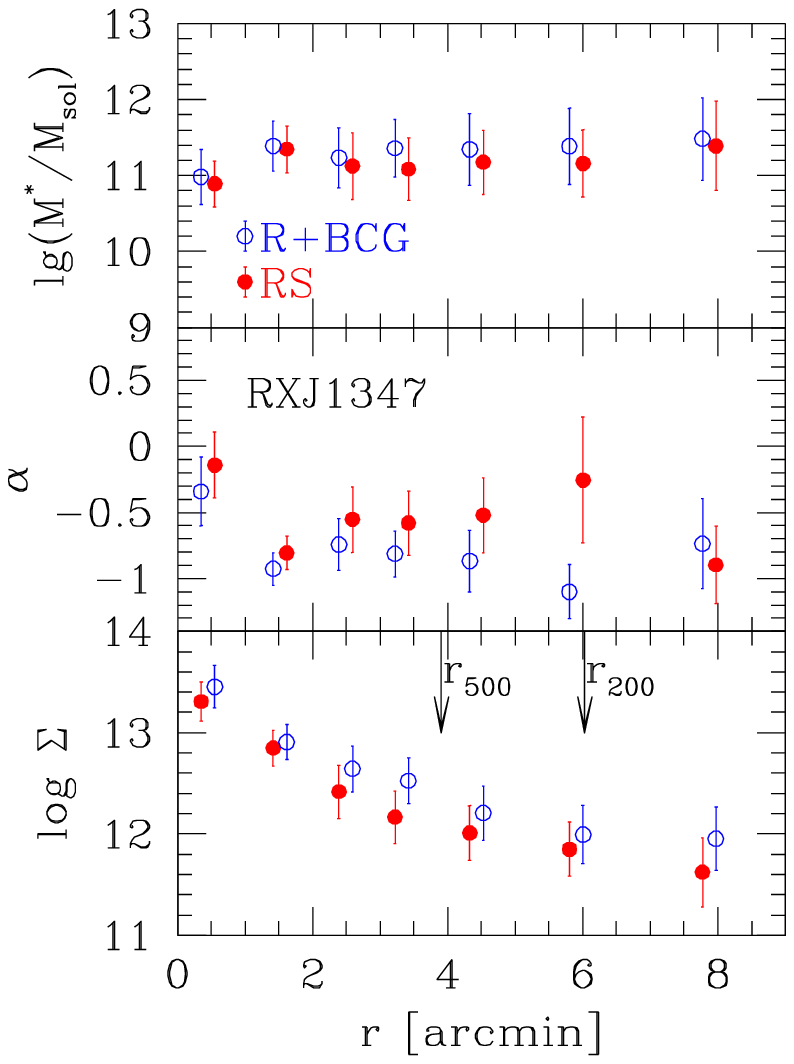,width=6truecm,clip=}
}
\caption[h]{Radial profile of the characteristic stellar mass
(upper panels), faint--end slope $\alpha$ (central panels) and
projected surface stellar mass density (bottom panels) of the
three clusters (MACSJ1206, MACSJ0329, and RXJ1347, from left to right).
The central--left panel report the 
fitted linear trend.  Arrows indicate $r_{\Delta}$ values and are from
Merten et al. (2014). The surface stellar mass 
density $\Sigma$ is in $M_{\odot}/$Mpc$^2$
units. 
}
\end{figure*}

To reduce the background contamination, we 
remove galaxies with $V-z'>2(V-R-1)+1.7$ mag. As 
checked with CLASH HST 16-band photometric redshifts (these are available 
for galaxies in  the inner arcmin only) these are in the
cluster background with no exception and are recognized as such
because they have the 4000 \AA \ break
in the $R$-band filter, not in the $V$ band as do cluster members.
Residual background galaxies are statistically accounted for,
as mentioned below. Using MACS1206 spectroscopic redshifts 
(Biviano et al. 2013), we found that less than 0.5\% of the mass 
is missed because of the  color cut adopted.

We adopt radial bins equal to those adopted in Umetsu et al. (2014),
after merging three Umetsu et al. (2014) bins into one, in order
to have the same radial bins in the 
computation of the radial
stellar--to--total matter ratio profile. Specifically,
we adopt the following radii: $0, 0.9, 2.1, 2.8, 3.8, 5.1, 6.7,$
and, except for MACSJ1206, $9$ arcmin as extremes of the considered
coronae.

To correct for the missed flux (and hence mass) 
coming from the galaxy outer regions 
(e.g., Blanton et al. 2001; Andreon 2002a,b) and the intracluster light 
we increased the measured flux of all radial bins by 15\% 
(Blanton et al. 2001; Andreon 2002a,b, 2010, 2012; Bahcall \& Kulier
2014).

\section{Characteristic mass, faint--end slope, and surface density stellar
mass profile}

\subsection{Methods}

To derive the galaxy mass function, we adopted a Bayesian approach,
as has been done for the luminosity function of 
other clusters (e.g., Andreon 2006, 2010, 2012, Andreon et al. 
2006, 2008, Meyers et al. 2012), and we account for the residual background
(galaxies in the cluster line of sight not  removed by
the color cut), which is estimated  
outside a radius of 3 Mpc and inside the same Subaru pointing
with  the cluster, hence fully guaranteeing 
homogeneous data for cluster and control field. 
We adopted a Schechter (1976) mass function for cluster galaxies and
we modeled the background counts with a power law of degree $3$ (i.e., with
four free parameters). 
We computed the total mass in stars from the integral, over all masses, of mass times the
galaxy mass function. 

The likelihood expression 
is derived and given in  
Andreon, Punzi \& Grado (2005), which is an
extension of the Sandage, Tammann \& Yahil (1979) likelihood expression  
for the case where a background is present.
It allows signal fluctuations,
i.e., it allows  the model and observed number of galaxies 
to differ, accounting for 
fluctuations in the number of galaxies. 
Because of these number fluctuations,
the stellar mass (and luminosity) inside
a radial bin has a uncertainty larger than just the one induced by
the finite number
of collected photons even in ideal conditions, such 
as in the absence of a background and with perfect knowledge of the luminosity to
mass conversion.

In the mass function computation, attention should be paid 
to the BCG, which
might not be drawn from the Schechter (1976) function 
(e.g., Tremaine \& Richstone 1977). 
This situation occurs 
for the inner corona of RXJ1347 and MACSJ0326, in which case  
we fit all galaxies excluding the two BCGs, and add their mass  
contribution separately. 

We adopt weak, uniform priors
on all parameters, thus allowing the cluster parameters to be
radial--dependent (i.e., allowing mass segregation, a different contribution
from low-mass galaxies, etc.) and we fit the individual galaxy data (i.e.,
we do not bin in mass). 
We checked convergence of the Markov chain Monte Carlo by running multiple chains,
and by inspecting their traces, their autocorrelation functions, and
how well the model fit the data. 
The Bayesian approach allows us to easily propagate
uncertainties and their covariance into derived quantities, such as
the surface stellar mass density. 
For example, we do
not assume a perfect knowledge of the background contamination
and instead marginalize over its uncertainty, allowing us to propagate
this uncertainty into derived quantities. As a visual check and for
illustration, we also
computed the mass function by binning galaxies in mass bins,
as is done for the luminosity functions by 
Zwicky (1957), Oemler (1974), and many papers since then. 

Our analysis measures the projected stellar mass, i.e.,
within a cylinder of nonnegligible width in the line of sight.
While our measurements are projected
on the line of sight (i.e., 2D, not 3D)  they are differential in the
plane of the sky (i.e., we measure $f(r)$, not $f(<r)$).
Weak lensing masses measure projected masses including 
contamination by large-scale structures (LSS)
located near  the cluster redshift, as our
stellar masses do.
This is  advantageous when we compute the stellar--to--total matter
ratio  
profile because
we are counting masses in similar volumes for both the numerator and
denominator.

\subsection{Results}

In this section we present the mass function of the three clusters; we show 
that the characteristic mass $M^*$ is
fairly constant as a function of clustercentric radius and
across the three clusters, that the faint--end slope of the mass function
shows a radial dependence indicating mass segregation, and we derive
the surface stellar mass density profile of the three clusters.

Figure~1 shows the galaxy mass function of RS and R+BCG galaxies in MACSJ0216 with
$\log M/M_\odot >9.7$, our mass completeness limit, at various clustercentric
radii. The very deep Suprime--Cam data very tightly constrain the cluster
mass function. 
The three--parameter Schechter function accurately describes the behavior of
the explored 2.5 dex in galaxy mass.
Appendix A shows the mass function of the other two studied clusters.

The top and middle panels of Fig.~2 shows the radial trend of the derived
characteristic mass $M^*$ and of 
the faint--end slope $\alpha$ of the Schechter (1976) mass function 
of the three studied clusters. There is no evidence for
a radial--dependent  characteristic mass, nor for a widely 
different value that depends on the considered galaxy population, nor for
a difference from cluster to cluster: 
$\log M^*/M_{\odot}=11.27\pm0.04$ (error on
the mean) for red--sequence galaxies, and $\log M^*/M_{\odot}=11.42\pm0.03$
for R+BCG galaxies. The radial independence of the massive
end of the mass function suggests that environment plays a minor
role in shaping the galaxy mass of the most massive galaxies. This
interpretation, however, assumes that the current 
environment is still closely correlated to the environment where the galaxy mass 
was shaped, a correlation that we know is 
largely masked by the backsplash population
(e.g., Gill et al. 2005).

The faint--end slope of MACSJ1206 steepens as we move toward
larger radii. For R+BCG the slope of the $\alpha$ vs $r$ relation is
$-0.101\pm 0.028$ (the slope is $<0$ with 99.97 \% probability). 
While for the R+BCG population the radial trend may be due
to a change in the relative proportion of RS and non--RS galaxies, the
effect is also significant for the RS population only, indicating
a mass segregation. 
The radial trend of the faint--end slope of the mass function
of the other clusters
is less clear. Although there are various hints of possible 
cluster--center dependencies of the faint--end slope (perhaps easier
to spot in the mass functions shown in Appendix A), none of them
is, alone, overwhelmingly statistically significant.

The difference between the regular behavior of the $\alpha$ profile
of MACSJ1206 and the apparently erratic behavior of the other two clusters
is striking and suspicious. To investigate if the profiles were contaminated by another
localized structure (a group, or another cluster, on a close line of sight) not
recognized as such, we repeated the analysis but
dividing the cluster area in octants and removing the two octants
richer in galaxies (to remove the possible contaminating 
structure) and the two poorest ones (not to bias the average). We found
indistinguishable profiles (apart for the reduced S/N), indicating
that our profiles are robust to contaminating clusters/groups.
Inspection of the 2D
weak--lensing mass maps of Umetsu et al. (2014) shows substructures
in the mass distribution of MACJ0329, which has an obvious secondary 
mass peak also pointed out by the shape of the gravitational arc and
the distribution of galaxies. RXJ1347 also shows a secondary mass
peak, also pointed out by
the two similarly bright BCGs. MACSJ1206 instead shows a more regular 
mass map and therefore could be in a more evolved dynamical status of the
other two clusters. It seems, therefore, that 
mass segregation (i.e., the regular steepening of the faint--end slope) 
is most obvious in the most evolved cluster, MACSJ1206. 
A larger sample of clusters is
however needed to secure such an interpretation.

The bottom panel  of Figure~2 shows the projected surface mass density
$\Sigma$ 
(i.e., the mass included in the corona divided by its area) of both
galaxy populations. These are the main results of this section, and one
of the two main ingredients of the computation
of the stellar--to--total matter ratio described in
the next section.

Fitting the  stellar mass profiles with a NFW radial profile, for
example to measure the relative distribution of
stellar and total mass following previous works, is a risky
operation because we known that the NFW profile
is supposed to represent
the mean, not every single, halo. Many clusters are known to
possess total and stellar mass substructures and to live in complex
environments. Because the model radial profile does not
capture the complexity of the data being fitted,
the best fit parameters depend, for example,
on the fitted radial range. As mentioned, 
MACJ0329 and RXJ1347  both possess a secondary massive halo. 
Close inspection of the total surface mass density profile in 
Umetsu et al. (2014) shows deviations from a NFW radial profile, 
and indeed these are also present in our stellar mass density profiles. 
Therefore, we avoid fitting NFW radial profiles to stellar mass
density profiles to measure the relative distribution of
stellar and total mass.

To summarize, the characteristic mass is radial and cluster independent 
and shows minor differences depending on the considered population. 
The faint--end slope is, instead, radial
dependent. In the 
more evolved cluster, MACSJ1206, the faint--end regularly steepens with
increasing clustercentric radius. The other two clusters, both
characterized by substructure, show more erratic faint--end slope radial
profiles. Irrespective of the regular or erratic behavior of
the radial profile of the faint--end slope, 
 we surely cannot force the contribution
of low-mass galaxies to be radial--independent, 
a modeling first enabled in this work.

\section{Radial profile of the stellar--to--total matter ratio}

In this section we investigate whether stellar mass and total
mass are similarly spatially distributed by computing
the radial profile of the
stellar--to--total matter ratio. 
The two most important improvements of this work are the use
of a radial--model--free estimate of the total and stellar mass,
and the building of a framework able to properly use these data.
We must avoid 
operating (rebinning or dividing) on the data, especially
the weak lensing data, because this would 
destroy or distort the signal and noise  there. In fact,
there is an
important covariance that naturally arises between weak lensing mass estimates in different
coronae  because  they satisfy the observed
cumulative mass constraint (Coe et al. 2010). 
One of the coronae
in which stellar mass is measured consists of the union of three
Umetsu et al. (2014) radial bins and weak lensing estimates
extend to clustercentric radii larger than our
stellar mass determination. Radial bins
cannot be dropped or ignored (e.g., because they lack the corresponding
stellar mass estimate) and we have to
operate on the model, not on the data.  The adoption of 
a Bayesian approach (Gelman et al. 2003) naturally solves
all these problems.

\subsection{Methods}

Umetsu et al. (2014) give the posterior distribution of the 
surface total
mass profile in 11 radial bins in the form of a maximum likelihood estimate
$s_i$ and its covariance matrix $C^{stat}$; this covariance matrix is a measure
of the uncertainty, inclusive of the covariance across radial bins,
of the mass estimate as derived from the joint likelihood analysis 
of shear and magnification data,
\begin{equation}
m \sim \mathcal{N}_{11} (s,C^{stat}) \quad ,
\end{equation}
where $\mathcal{N}_{11}$ is the multinormal distribution in $11$ dimensions
(the 11 radial bins) and the symbol $\sim$ reads ``drawn from''. 

Projection noise (corresponding to $C^{lss}$ in Umetsu et al. 2014
notation) similarly affects both weak lensing and our stellar
mass estimate and thus can be (and is) ignored.
Systematics (corresponding to $C^{sys}$ in Umetsu et al. 2014
notation) are accounted for by allowing the mass, $m_t$, to 
fluctuate around the true mass, $m$, by an amount given by $C^{sys}$:
\begin{equation}
m_t \sim \mathcal{N}(m,C^{sys}) \quad .
\end{equation}

\begin{figure*}
\centerline{
\psfig{figure=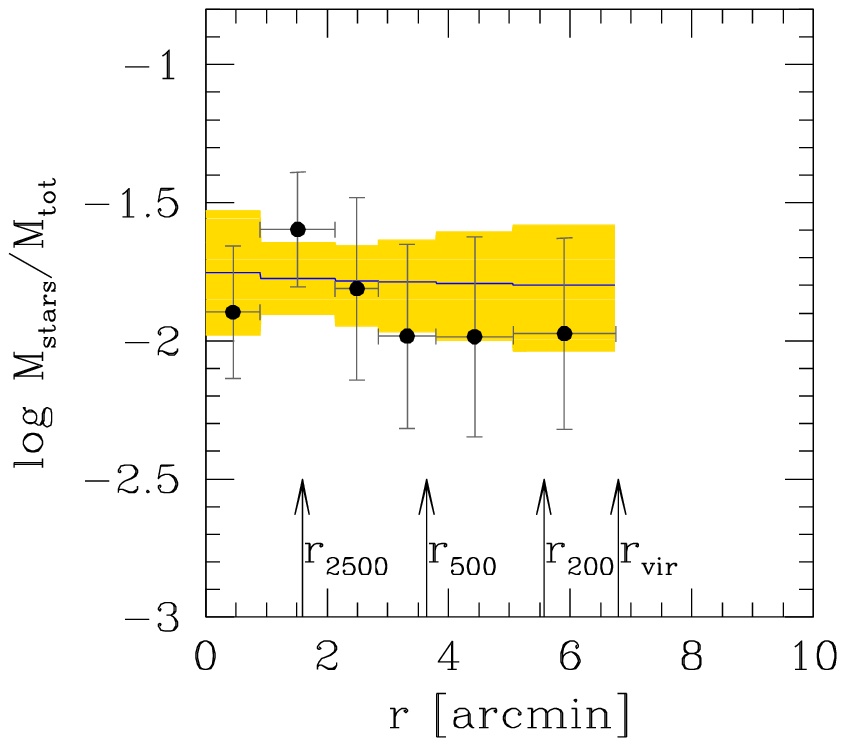,width=5truecm,clip=}
\psfig{figure=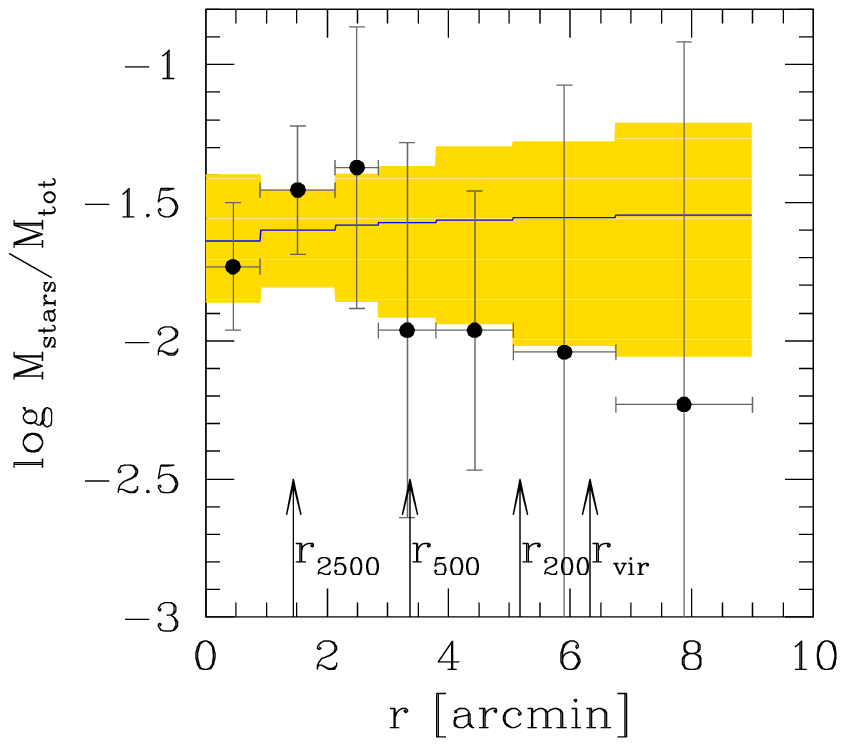,width=5truecm,clip=}
\psfig{figure=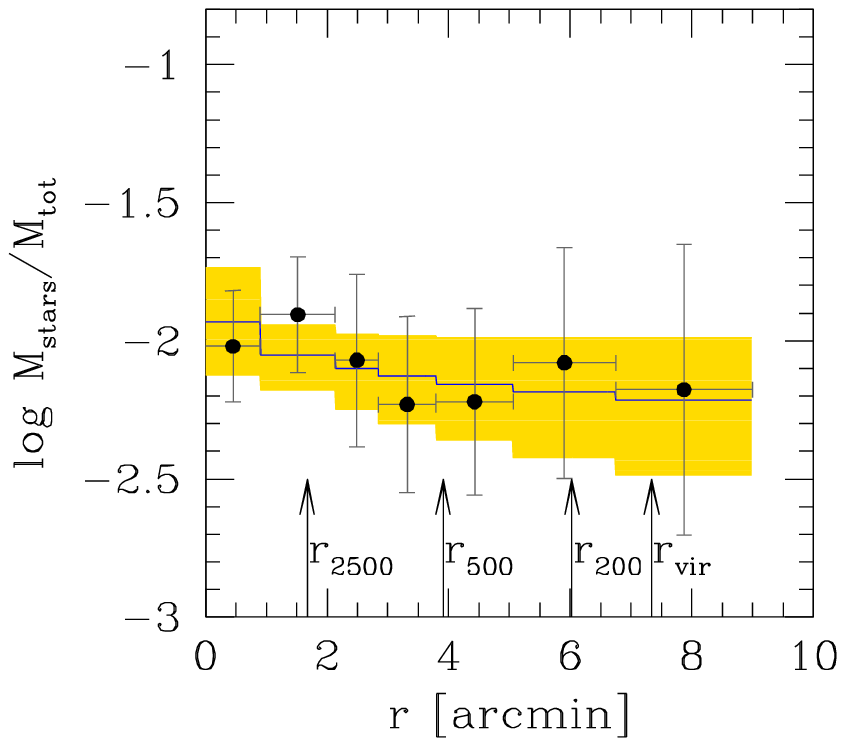,width=5truecm,clip=}
}
\centerline{
\psfig{figure=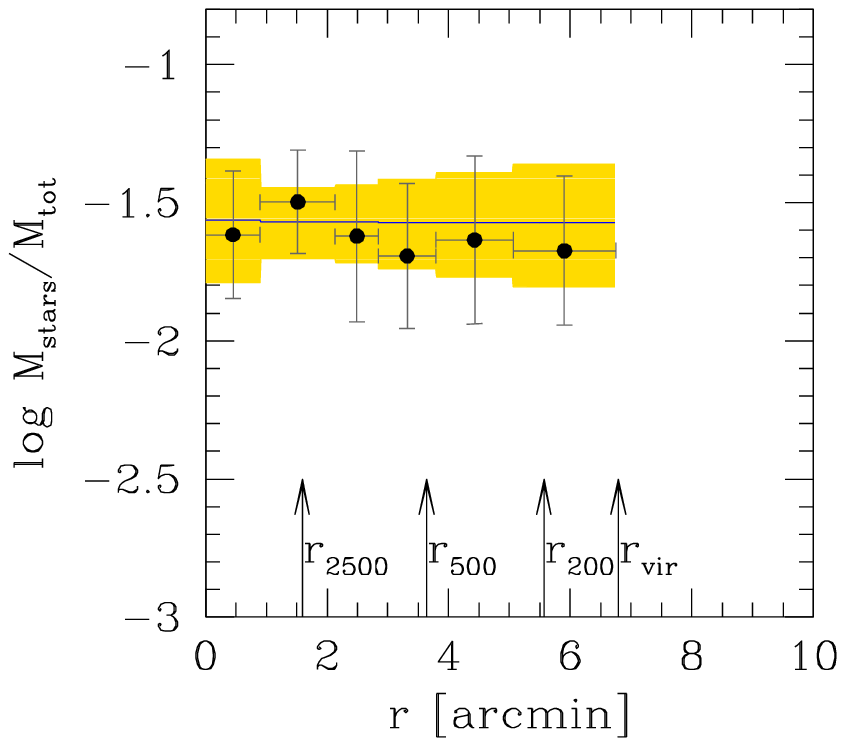,width=5truecm,clip=}
\psfig{figure=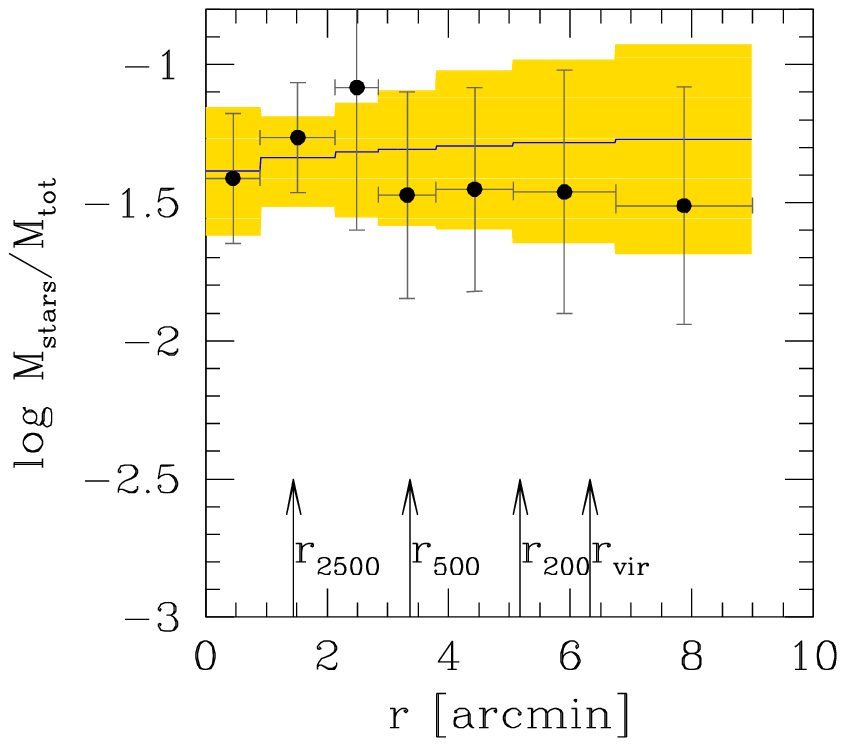,width=5truecm,clip=}
\psfig{figure=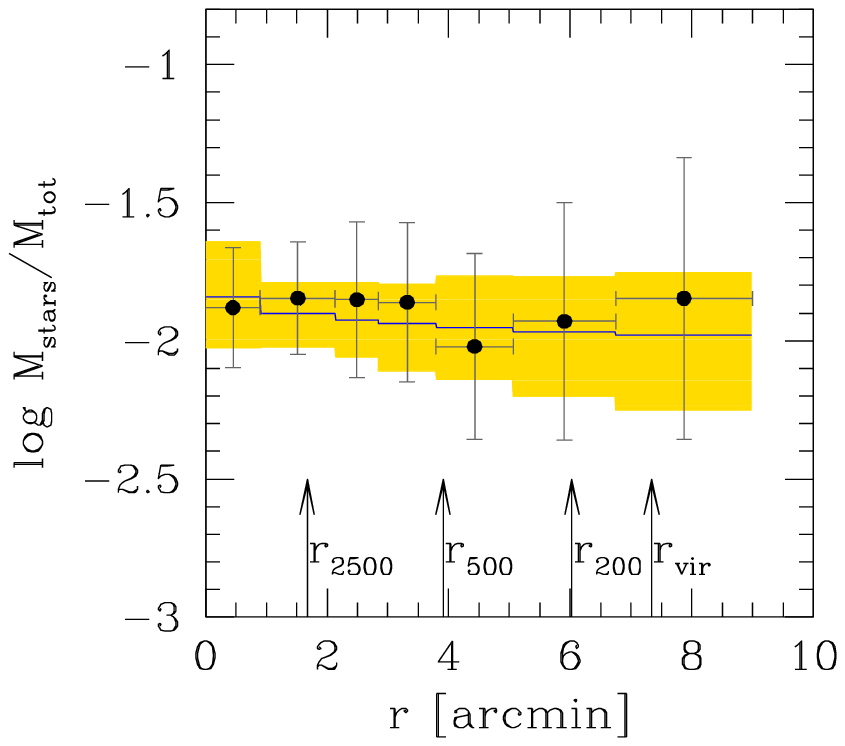,width=5truecm,clip=}
}
\caption[h]{Radial profile of the stellar--to--total matter ratio of
MACSJ1206 (left), MACSJ0329 (center), and RXJ1347 (right).
The upper panels only consider red sequence galaxies, whereas
lower panels consider red galaxies and BCGs. The multi--step function and
the shading indicate soundly derived values (posterior mean)
and 68 \% errors (highest posterior interval), whereas points
and error bars are approximate (see text).}
\end{figure*}

As in Umetsu et al. (2014), a positively defined mass $m_t$ is enforced in all
radial bins. We require that a $>10^{12}$ $M_{\odot}$ mass be present
in every radial bin and we checked that results do not depend on the
minimal mass we assumed.

Coming to the surface stellar mass profile, 
the results of our analysis can be summarized as
\begin{equation}
\log \Sigma \sim \mathcal{N}_l(\Sigma_m,C^{\Sigma}) \quad ,
\end{equation}
where $l$ is $6$ or $7$ (the number of stellar mass radial bins),
$\Sigma_m$ is the posterior mean (i.e., the point estimate in previous section), and 
the covariance vector $C^{\Sigma}$ is a diagonal matrix having as
elements the square of the posterior standard deviation (errors 
of previous section).
For our stellar
masses there is almost
no covariance across radial bins because the measurement
is local, considering the luminosity of galaxies inside the corona.
Instead, measurements are
global in weak--lensing estimates. In the stellar mass measurements 
the only source of covariance
comes from the uncertainty of the mean background 
stellar mass because it is inferred from a reference line of sight 
common to all radial bins. However, since the \emph{mean}
value is fairly well determined, the covariance induced by uncertainty
on the mean value is negligible and neglected (while fluctuations around
it are important and accounted for). In cases in which covariance is
also present  for stellar masses, Eq.~3 still holds, but the $C^{\Sigma}$
matrix is not diagonal.

To deal with our second radial bin, formed by the union of the 
second to fourth Umetsu et al. (2014) bins, we only need to 
remember that the mean mass surface density inside any combination
of bins is just the average over the concerned bins.

We assume a linear relation with slope $\gamma$ and intercept $\beta$
between the $\log$ of the stellar mass fraction and $\log$ of radius: 
\begin{equation}
\log \Sigma - \log(m_t) = \gamma \log (r/4)+\beta \quad .
\end{equation}
As in  Andreon et al. 
(2006, 2010), for example,  we adopt a $\log$
scale because fractions are positively defined, while a linear
scale would allow, at least in principle, unphysical (negative)
values for the fraction. We zero--point radii at $4$ arcmin
in order to have intercepts (stellar mass fractions) at radii
close to $r_{500}$.

Finally, we adopt weak priors on the slope $\gamma$ and intercept $\beta$, 
respectively a Student--$t$ distribution (which is uniform on the angle),
as in Andreon \& Hurn (2010), and a Gaussian with large $\sigma$.
These  equations are solved by    using a Marcov chain
Monte Carlo sampler (JAGS, Plummer 2010). For drawing
figures we instead operate on data and sum errors in quadrature and
ignore covariance across radial bins.

The method detailed above, together with the methods described in Sect.~3.1,
is the second result of this work, allowing us to
confidently state the significance of found radial trends, or lack 
of them.

\subsection{Results}

Figure 3 shows the derived radial profiles  of the stellar--to--total matter 
ratio.

RXJ1347 has a stellar fraction lower 
than the other two clusters, which is unsurprising because
the stellar mass fraction decreases with cluster mass 
(Andreon 2010, 2012; Gonzalez et al. 2014), 
and this cluster has a much larger mass than the other two clusters (which
are of comparable mass).

The radial profiles are very flat, especially the properly 
derived values (multistep function), with no evidence of gradients.
Indeed, $|\gamma|<0.1$ with error $0.33-0.50$ (depending on
the sample). This is one of the two main results of this work and
shows that cold baryons (stars) are distributed as total mass
over a wide range of clustercentric radii, $ 0.2 \lesssim r/r_{200}
\lesssim 1.6$. 

From a galaxy evolution perspective, 
the flatness of the stellar--to--total matter
ratio is hard to interpret. On the one hand, it is
somewhat expected because galaxy harassment (Moore et al. 1996) 
and the cluster tidal field should not affect
the stellar mass given that the latter is locked deep in the potential 
wells of galaxies.
On the other hand, harassment and tidal field
are able to strip part of the dark matter of both
individual galaxies and infalling groups; i.e., they increase
the stellar--to--total matter ratio.
This infalling material, in turn, has a
larger stellar--to--total matter 
ratio according to observations (e.g., Andreon 2010, which is also 
a consequence of the different shape of the luminosity and mass functions,
Monster et al. 2010) and this, together with the stripping of part of the fragile dark halo, 
increases the contrast between the high ratio of
infalling material and the low
ratio of already fallen one.
To keep the cluster (=fallen+infalling) profile flat, a
redistribution of the dark matter component is therefore needed. Given
that only a few crossing times have
taken place in a Hubble time because of the large cluster size,
the redistribution have not to take several crossing times
to keep the cluster profile flat. Therefore, a 
mechanism much faster than the crossing time, and some fine-tuning, is needed to
keep flat the ratio radial profile. 

Neither a fast mechanism nor  fine-tuning is
needed if the infalling material includes not only halos with large
values of stellar--to--total matter (i.e., groups and massive galaxies), but 
also halos of much lower mass (less massive galaxies and 
perhaps halos without any galaxy
inside it), because the latter  have a stellar content per unit mass comparable with, or
lower than, the cluster composition (Hudson et al. 2015). 
A flat radial profile can be
therefore naturally achieved if the stellar--to--total matter
ratio of each infalling shell has a value for the
ratio close to the already fallen material. 
Indeed, the $M/L$ also seems nearly constant at the largest
scales (Bahcall \& Kulier 2014, and references therein), making
fine--tuning and a fast mechanism
unnecessary. However, the latter observations are measuring something
conceptually different from a radial profile: measurements
at different scales pertain to different objects, not to the
same objects observed at larger radii. This means that our proposal 
still needs to be observationally confirmed.

It would be interesting to quantify these ideas using numerical simulations
of the growth of clusters from their surrounding environments;
however, this requires a numerical simulation of a quality higher than
currently available. To test these ideas we need simulations
with halos having a mass--dependent stellar--to--total matter ratio
and current simulations fail to reproduce this feature (Andreon 2010,
Young et al. 2011; Planelles et al. 2013; Battaglia et al. 2013).

From a cosmological perspective,
the close resemblance of total and stellar matter profiles can be
useful
to improve lensing-- and dynamical-based mass estimates by adopting the
stellar matter profile as a prior for the total matter profile. The
common flat $M/L$ radial profiles found in the literature (see Introduction
and next section) 
support our results obtained from a more detailed analysis (of only
three clusters, however).
This prior has
already been exploited in some strong--lensing analyses such as Grillo et al.
(2014) and Kneib et al. (1996), and our result of a flat radial
profile supports their use. 
The similarity of total and stellar matter profiles is also
cosmologically attractive for using
the stellar matter profile as a mass proxy. This can be
achieved either using the profile
alone (i.e., only using the cluster size), 
or in combination with
the stellar matter fraction. The simplest
implementation of the former approach needs
to assume a parametric form for the radial profile (e.g., a NFW 
profile), although nonparametric forms for the radial
profiles are possible too, and have been used since Kneib et al. (1996).
Instead, using the stellar matter profile together with the stellar matter 
fraction requires, at first sight, knowledge of mass (to compute the fraction).
However, this dependency (degeneracy) may be broken as is done
for the $Y_X$ and $Y_{SZ}$ proxies (which requires knowledge 
of $r_{\Delta}$, i.e., $M_{\Delta}$) by assuming the observed scaling relation
(the mass--stellar fraction
relation in our case). We leave this  exploration to a later paper.

Finally, the flat stellar--to--total matter ratio profile is a
technically important ingredient whose knowledge is needed for 
fitting the trend
between stellar mass fraction and halo mass: if the radial profile is
flat, the uncertainty on $r_{500}$ induces a complete covariance between
total and stellar masses.

\subsubsection{Comparison with previous works}

Before comparing our results to literature works, we emphasize
differences between related, but distinct, concepts:
light cannot be interpreted as
stellar mass if the age of the stellar population is 
unconstrained, which makes the $M/L$  
different from the stellar--to--total matter ratio. Furthermore,
as already mentioned, the radial profile of the stellar--to--total matter 
ratio is also conceptually different from 
the increase of the mass--to--light
ratio with spatial scale (studied in, e.g., Bahcall 
et at. 1995 and references therein).
Finally, 
one may
have a constant stellar--to--total mass fraction and a luminosity segregation 
(brighter galaxies more concentrated in the cluster center, as in MACSJ1206
for example).
Therefore the presence (or absence) of a luminosity segregation is
not a measure of the stellar--to--total fraction profile, in particular of
its radial trend.

Our flat radial profiles in stellar--to--total matter ratio confirm 
the many flat $M/L$ profiles found in previous
works and improve on them because they are based on state--of--the--art weak lensing.
Unlike most previous works, our work offers a better sensitivity
to radial gradients because it uses differential quantities (most previous
works compute $M/L(<r)$, not $M/L(r)$),  does not assume a parametric
form for the mass profile,  does not miss stellar mass
in non--RS galaxies or in galaxies of low--mass, 
 discriminates star formation from stellar mass,  does not assume a constant
faint--end slope, and presents a sounder analysis accounting for
error and covariance terms ignored in past works.
Compared to analyses based on an ``ensemble'' cluster, such as the
dynamical analysis by Carlberg, Yee \& Ellingson (1997), our result
is independent of the assumption of the dynamical equilibrium and
orbit distribution and, being derived on individual clusters, ensures
that the found constancy of the stellar--to--total matter 
ratio is not an artifact of the scaling laws adopted to construct
the ensemble. 

Unlike a  few literature findings, we do not
find decreasing stellar--to--total matter ratio profiles. 
In particular, we do not confirm the tentative 
trend seen for RXJ1347 by Medezinski et al. (2010)
which is remarkable because both weak lensing and stellar masses are
based on non--independent data (Suprime--Cam images). 
We note, however, that Medezinski 
et al. (2010) use parametric weak--lensing mass estimates (from
Broadhurst et al. 2008), not the radial nonparametric estimate
we use (from Umetsu et al. 2014), and we have already remarked that
individual clusters show deviation from the universal radial
profile, including RXJ1347. We also warn that Medezinski 
et al. (2010) use very uncertain corrections for the background 
contamination, so uncertain that they are outside the physical
range (0 to 100 \%) half of the time. We also do not  confirm
the main Annunziatella et al. (2014) conclusion
on MACSJ1206, largely based on the same Subaru data used here,
claiming a more concentrated dark matter distribution than stellar
mass. Their conclusion, just as is the Medezinski et al. (2010) conclusion, 
is conditional upon the assumption
that MACSJ1206 has a NFW profile. It is notable in this context
that Young et al. (2014), having collected various evidence for
the presence of a second component aligned in the cluster line of sight, 
fit a multi-halo NFW profile to
weak lensing data of this cluster. When Annunziatella et al. (2014) 
get rid of this assumption (their Fig.~12)
the ratio of the two mass profiles shows no trend with radius
and is consistent with a single constant at all radii, precisely
as our radial stellar--to--total matter ratio profile is. Finally, although our data
are sensitive to gradients as large as those observed in Abell
576 by Rines et al. (2001), we do not observe them in
the studied clusters. As remarked in Rines et al. (2001), the statistical
significance of their found gradient is hard to establish because
their analysis does not account for the mass covariance across
radial bins; i.e., it lacks the necessary  formalism introduced  
in our Sects.~3.1 and 4.1
several years after their paper. This shows, once more, that 
the significance
of a claimed radial term, and the upper limit in case of a
lack of detection, depends on which terms are included
in the error budget, and, generally speaking, 
other works have smaller errors
in spite of worse data because 
they include an inferior number of terms in the error budget
(see Appendix B for details). 

Finally, we conclude by pointing out the main limitation of our work.
Since we use radial bins (largely fixed by
weak--lensing measurements), we cannot explore the 
trend of the stellar--to--total matter ratio at scales lower than half 
the inner bin size, i.e., within the inner 150 kpc or $0.15 r_{200}$. Measurements
performed by Newman et al. (2013) nicely complement our work.

\section{Summary}

In this paper we used the extremely deep 
multicolor wide--field Suprime--Cam optical images and 
state--of--the--art accurate, yet projected,  
weak lensing masses to derive the relative
distribution of stellar and total matter of three massive 
intermediate--redshift clusters. Weak lensing does not oblige
us to make assumptions about the cluster dynamical (or hydrostatic)
status, and our choice of using nonparametric estimates
makes our derivation free
from assumptions about the shape of the mass radial profile.
Unlike most previous works, our analysis offers a better sensitivity
to radial gradients  because we use differential quantities (however
projected along the line of sight). Our analysis pays special
attention to items that are usually underrated: the stellar mass 
in galaxies outside the red sequence; those in galaxies of low mass,
in particular below the observational limit ($\log M/M_{\odot}=9.7$ in
our study); possible mass segregation and cluster substructure; 
uncertainty on the mean background value and
background fluctuations (in number and in mass at a given total number);
and discrimination of star formation from stellar mass. 
Given that we account for all source of errors (that we are
aware of), and also mass covariance across radial bins, 
we can confidently state the significance of found radial trends, or lack 
of them. Building the statistical framework able to properly use 
the data is one of the two main results of this work.

The characteristic mass is radial and cluster independent and shows
minimal differences depending on the considered population 
($\log M^*/M_{\odot}=11.27\pm0.04$ for red sequence galaxies  
and $\log M^*/M_{\odot}=11.42\pm0.03$
for R+BCG galaxies). The faint--end slope is, instead, radial
dependent, indicating mass segregation. 
In MACSJ1206, the cluster more regular and evolved and
showing less substructure, the faint--end regularly steepens with
increasing clustercentric radius. The other two clusters, both
characterized by a second mass clump, show more erratic radial behaviors.
Whether or not the radial profile of the faint--end slope is regular or 
erratic,  the faint--end slope surely cannot be forced to be radial independent 
in the total luminosity computation.

The main result of this work is that the radial profile of the stellar--to--total
mass is constant from about 150 kpc to the virial radius and beyond
for the three studied clusters. Our sensitivity to gradients is
a factor of two per decade in radius.
The constant fraction is due to the remarkable similarly
of the spatial distribution of stellar and dark (total) mass and
implies that dark matter and cold baryons (stars) are very tightly coupled
over a remarkable range of environmental densities going from the
cluster core to its virial radius.

\begin{acknowledgements}
I acknowledge Keiichi Umetsu for giving me the mass profiles
in tabular format and for useful discussion, the CLASH team for distributing the
reduced Subaru images, Susan Planelles and Nick Battaglia
for passing me in electronic form their stellar--to--total
profiles, Elinor Medezinski and Andrea Biviano for useful discussion, and
Kristin Riebe for help in using cosmological simulations. Annalisa Pillepich
is thanked for an attentive reading of this paper.

\end{acknowledgements}

{}

\appendix

\section{Other Figures}

This appendix presents the mass function of RS and of R+BCG galaxies in
RXJ1347 and in MACJ0329 at various clustercentric
radii.

\begin{figure*}
\centerline{
\psfig{figure=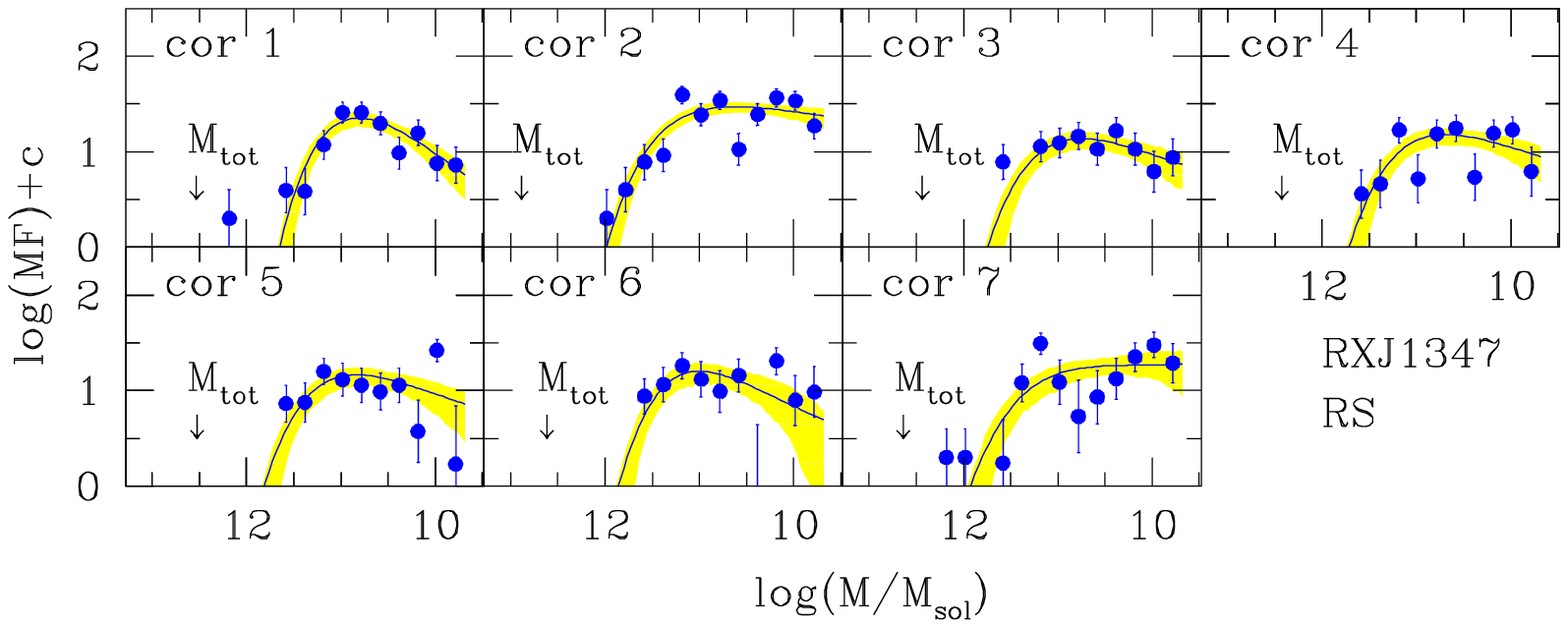,width=14truecm,clip=}
}
\centerline{
\psfig{figure=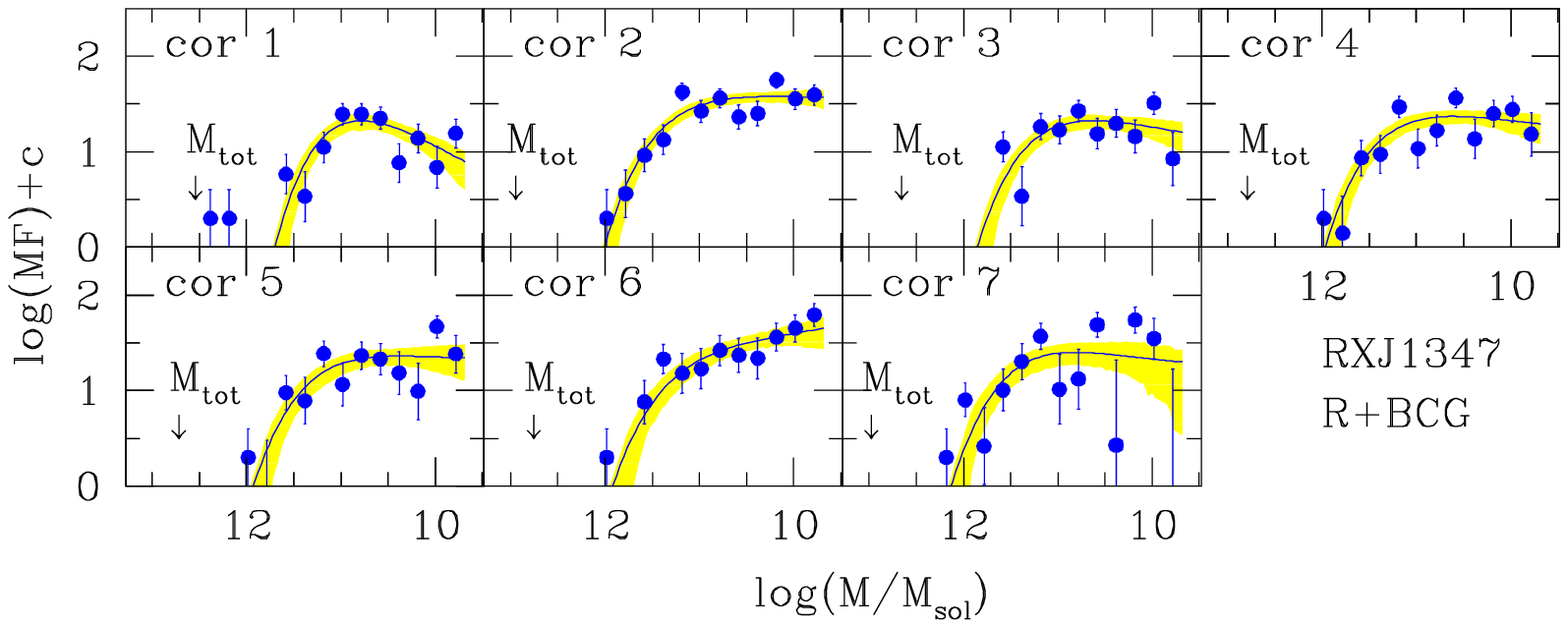,width=14truecm,clip=}
}
\caption[h]{Mass function of red sequence (upper panels) and 
red galaxies plus BCG (lower panels) of RXJ1347 in various 
clustercentric radial bins, as detailed in the panels. Caption as in Figure~1.
}
\end{figure*}

\begin{figure*}
\centerline{
\psfig{figure=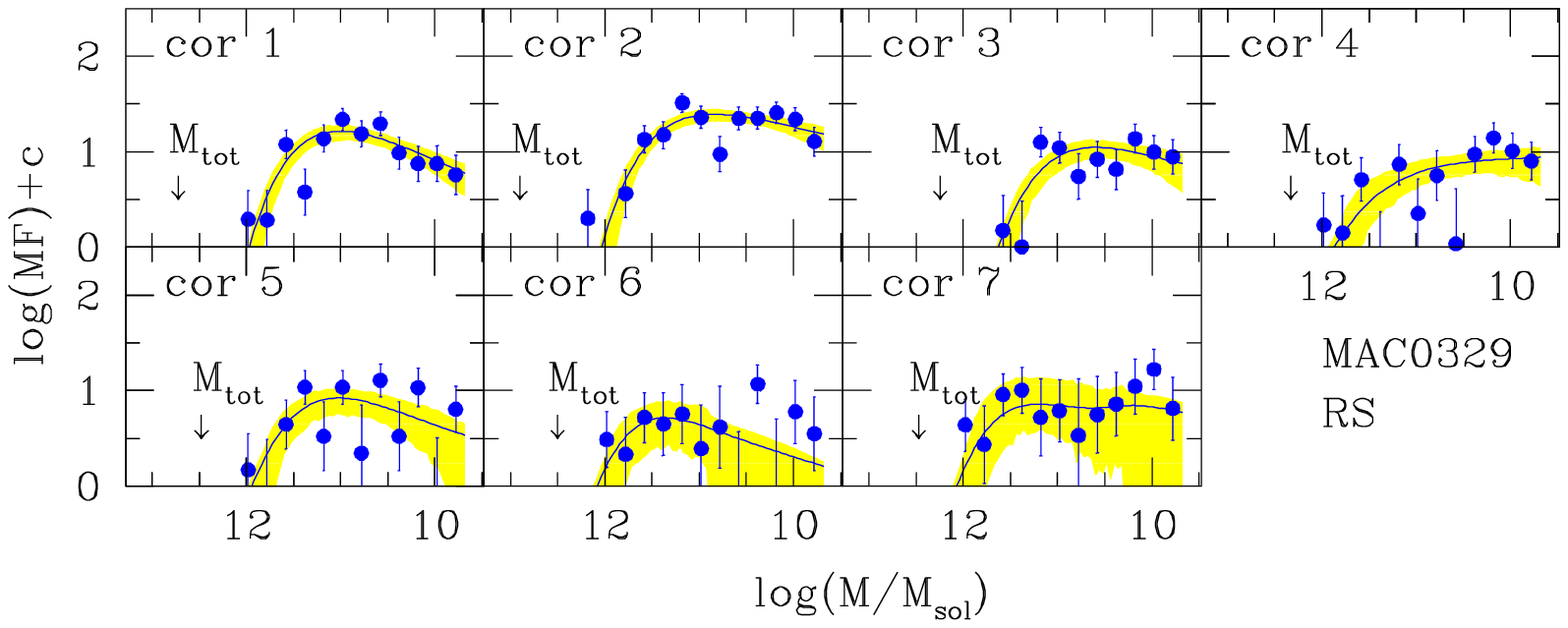,width=14truecm,clip=}
}
\centerline{
\psfig{figure=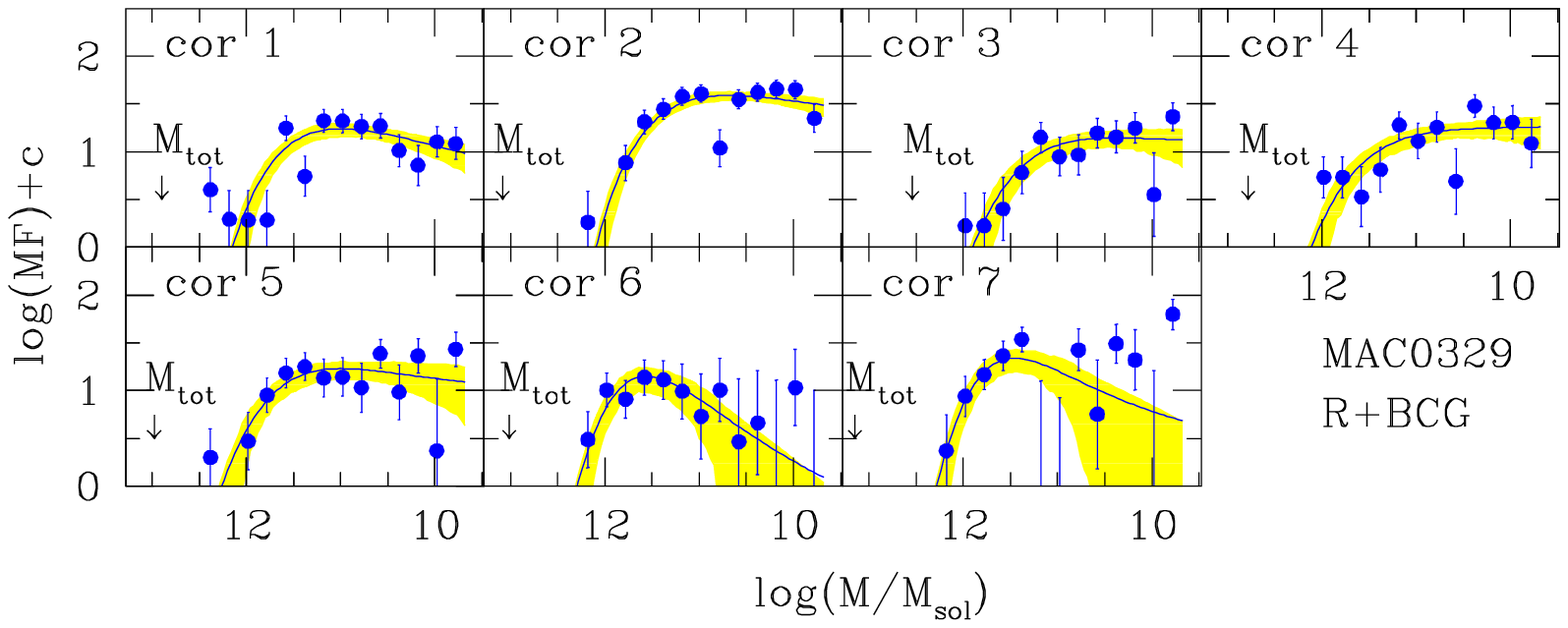,width=14truecm,clip=}
}
\caption[h]{As previous figure, but for MACSJ0329.
}
\end{figure*}

\section{Fitting details and systematics}

A proper estimation of the
error is paramount to properly claim as significant a stellar--to--total matter
(or $M/L$) ratio radial trend. 
In general, our stellar mass
errors are much larger than those returned by
procedures adopted in literature, sometime by a factor of 10 because
these works chose not to include in the error budget terms
that we instead consider important, for example
Poisson fluctuations of the total
number of background galaxies and of their number at a given mass. 
We see no reason why
the background population in the cluster and control field lines of sight 
should be exactly average in number and total mass. We verified  the
existence of these well--known fluctuations by 
splitting the reference line of sight in parts. 
These number and mass fluctuations make the stellar
mass in the background population subject to fluctuations much larger than 
those coming from
photometric errors alone (i.e., the stochastic mass errors). 
This is a large
source of uncertainty
at all clustercentric radii except at the cluster center, where the
background contamination is minimal. It is modeled in our work, but
rarely included
in previous works.
Second, in addition to these Poisson fluctuations,
the parameters that describe the mass distribution
of background galaxies are measured with a finite precision, and
we need to marginalize over this uncertainty too (i.e., over the uncertainty of
the parameters describing the average expected background
population). This marginalization is naturally implemented in our
Bayesian approach.
This error term is minor for our own analysis because a large solid angle observed
in fully homogeneous conditions is
available for the mean background estimation, but is major, yet
neglected, in some works.
Third, given the extreme depth of our data, the stellar
mass at masses lower than the mass completeness limit
is negligible for whatever faint--end slope $\alpha>-2$ for our study.
Nevertheless, we account for it by marginalizion.
Data of such depth are rare at best
in the literature, yet we are aware of no previous work 
computing the stellar--to--total matter (or $M/L$) ratio profile that 
marginalizes 
over the faint--end slope uncertainty. 
Indeed, most studies 
do not even account for the possible different faint--end slope at different
cluster--centric radii.

Some works use spectroscopic samples. Although
the background contamination can be derived using 
spectroscopic redshifts,
the sample with spectroscopic data is only a fraction of the whole sample
needed to measure the total stellar mass, 
and the background contribution has to be inferred for the subsample
without spectroscopic data. This is achieved by taking
the estimate derived from the spectroscopic sample, assuming that there are
no background fluctuations between the spectroscopic and photometric sample,
and that the derived correction is perfectly known, none of which
is true. Therefore,
and for the same reasons just discussed,
works using spectroscopic samples also 
have been too optimistic in their estimation of the 
stellar mass errors.

\end{document}